\documentclass[11pt]{article}

\usepackage[preprint]{acl}

\usepackage{times}
\usepackage{latexsym}

\usepackage[T1]{fontenc}

\usepackage[utf8]{inputenc}

\usepackage{microtype}

\usepackage{inconsolata}

\usepackage{graphicx}

\usepackage{amsmath}
\usepackage{booktabs}
\usepackage{multirow}
\usepackage{subcaption}
\usepackage{cleveref}
\usepackage{svg}

\usepackage{tcolorbox}
\usepackage{listings}

\newenvironment{fontppl}{\fontfamily{ppl}\selectfont}{\par} 

\newtcolorbox{promptbox}[2][]{
  floatplacement={#2},
  colframe=dark,colback=light!30!white,
  fonttitle=\small\ttfamily,
  fontupper=\small\ttfamily,
  title=#2,
  boxrule=0.5mm, 
  halign=flush left,
}

\definecolor{dark}{HTML}{064a6c}
\definecolor{light}{HTML}{efede1}

\title{Communication to Completion: Modeling Collaborative Workflows with Intelligent Multi-Agent Communication}

\author{
Yiming Lu$^{1,2}$\thanks{Work done during internship at Zoom}, 
Xun Wang$^{2}$, 
Simin Ma$^{2}$, 
Shujian Liu$^{2}$, 
Sathish Reddy Indurthi$^{2}$ \\
\bf Song Wang$^{2}$,
Haoyun Deng$^{2}$,
Fei Liu$^{1}$,
Kaiqiang Song$^{2}$ \\
$^{1}$Emory University \quad $^{2}$Zoom Video Communications \\
\texttt{\{yiming.lu,fei.liu\}@emory.edu} \\
\texttt{\{xun.wang,kaiqiang.song\}@zoom.us}
}

\begin{document}
\maketitle
\begin{abstract}
Multi-agent LLM systems have demonstrated impressive capabilities in complex collaborative tasks, yet most frameworks treat communication as instantaneous and free, overlooking a fundamental constraint in real world teamwork, collaboration cost. We propose a scalable framework implemented via \textbf{Communication to Completion (C2C)}, which explicitly models communication as a constrained resource with realistic temporal costs. We introduce the \textbf{Alignment Factor (AF)}, a dynamic metric inspired by Shared Mental Models, to quantify the link between task understanding and work efficiency. Through experiments on 15 software engineering workflows spanning three complexity tiers and team sizes from 5 to 17 agents, we demonstrate that cost-aware strategies achieve over 40\% higher efficiency compared to unconstrained interaction. Our analysis reveals emergent coordination patterns: agents naturally adopt manager centric hub-and-spoke topologies, strategically escalate from asynchronous to synchronous channels based on complexity, and prioritize high value help requests. These patterns remain consistent across multiple frontier models (GPT-5.2, Claude Sonnet 4.5, Gemini 2.5 Pro). This study moves beyond simple agent construction, offering a theoretical foundation for quantifying and optimizing the dynamics of collaboration in future digital workplaces.
\end{abstract}

\section{Introduction}
\label{sec:introduction}

\begin{figure}[h!]
    \centering
    \includegraphics[width=\columnwidth]{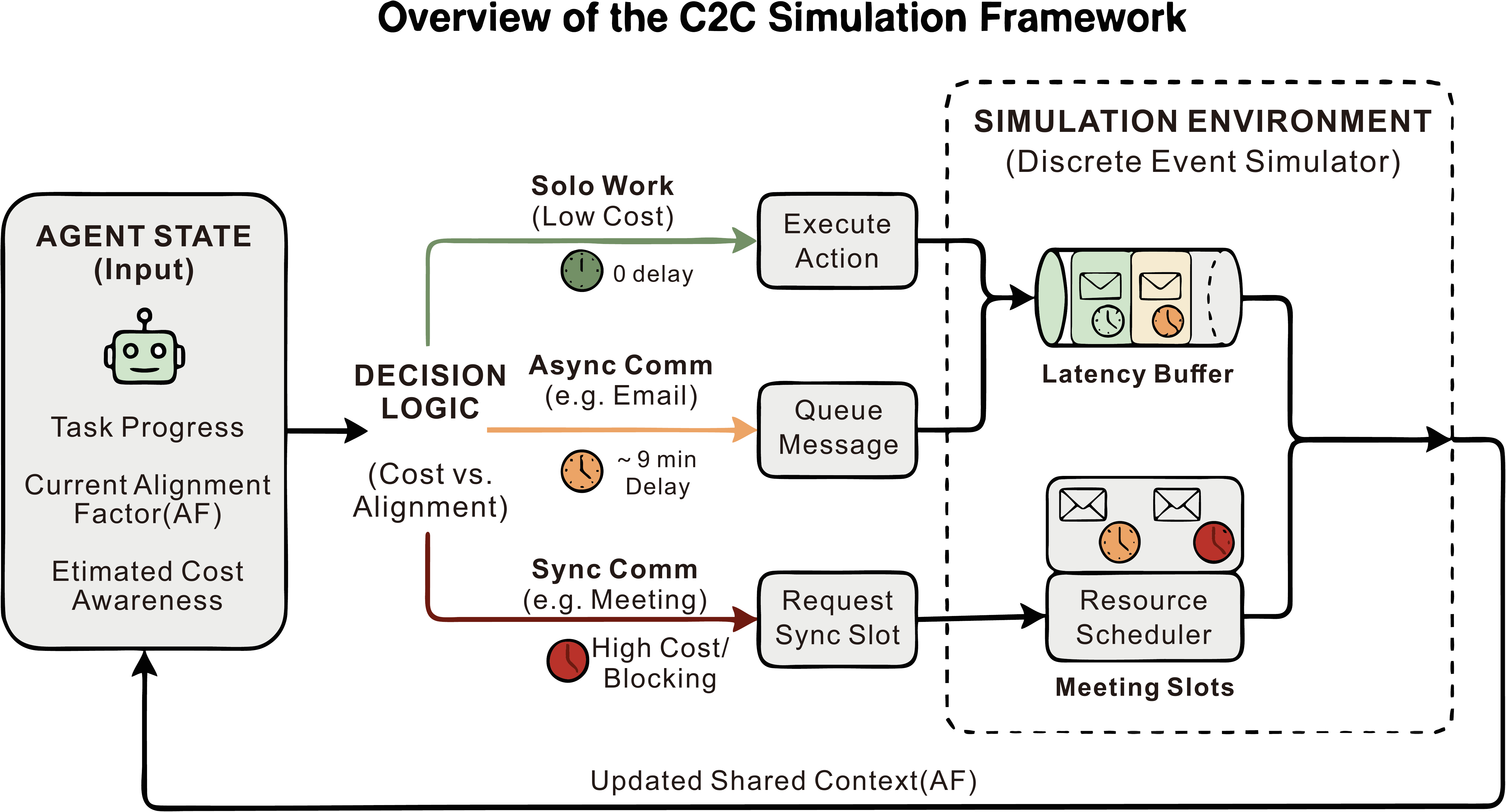}
    \caption{
    Overview of the C2C simulation framework. Agents autonomously select between execution and communication channels by weighing the Alignment Factor (AF) against modeled latencies, forming a closed-loop system where communication is treated as a constrained resource.
}
    \label{fig:my_svg}
\end{figure}

Effective teamwork in software development hinges on strategic communication. Teams must continuously balance the value of information sharing against its costs: synchronous meetings consume collective time, asynchronous messages introduce delays, and frequent interruptions fragment attention~\citep{mok2023challenging,yang2022effects}. Understanding \emph{when}, \emph{with whom} and \emph{how} to communicate remains a central challenge in organizational research~\citep{marlow2018does}.

However, systematically studying these dynamics in real teams is notoriously difficult due to the inability to control variables or replay scenarios. Recent advances in large language models (LLMs) offer a new approach to simulate workplace behaviors. Just as \citet{park2023generative} demonstrated that LLM agents can exhibit believable human-like social behavior in simulated communities, we propose using multi-agent LLM systems as a controlled simulation environment for investigating workplace collaboration patterns.

Despite this potential, current multi-agent frameworks are not well-suited for realistic workplace simulation. Systems such as AutoGen and MetaGPT typically abstract away communication costs, effectively assuming instantaneous information exchange~\citep{wu2024autogen,hong2024metagpt}. As a result, agents can exchange unlimited tokens without explicitly modeling temporal latency or cognitive costs (e.g., context switching) inherent in real workflows. Consequently, these frameworks cannot capture the fundamental trade-off between alignment and coordination cost, and may encourage excessive communication that would be unsustainable for human teams.

To bridge this gap, we introduce \textbf{Communication to Completion (C2C)}, a framework designed to simulate collaborative workflows under realistic constraints. C2C incorporates a discrete event simulation engine that explicitly models communication costs, enabling communication to be treated as a scarce resource to be optimized. Within this environment, we propose the \textbf{Alignment Factor (AF)}, a dynamic metric inspired by the concept of \emph{Shared Mental Models}~\citep{cannon1993shared}, to quantify how effectively agents build and maintain shared task understanding over time. We use AF both as an evaluation metric and as a control signal to guide when additional communication is worth its cost.

Our contributions are as follows:
\begin{itemize}
    \item We establish a computational sandbox for multi-agent collaboration that moves beyond instantaneous communication assumptions, enabling the study of agent behaviors under realistic temporal and cognitive constraints.
    \item We introduce the \textbf{Alignment Factor} as a quantifiable metric for shared task understanding. We show that optimizing for AF enables agents to dynamically balance the benefits of information sharing against the costs of coordination.
    \item Through experiments on complex software engineering tasks, we demonstrate that cost aware communication strategies outperform unconstrained interaction in efficiency. We further analyze emergent coordination patterns, showing how constraints induce more structured communication topologies.
\end{itemize}

\begin{figure*}[h!]
    \centering
    \includegraphics[width=1\textwidth]{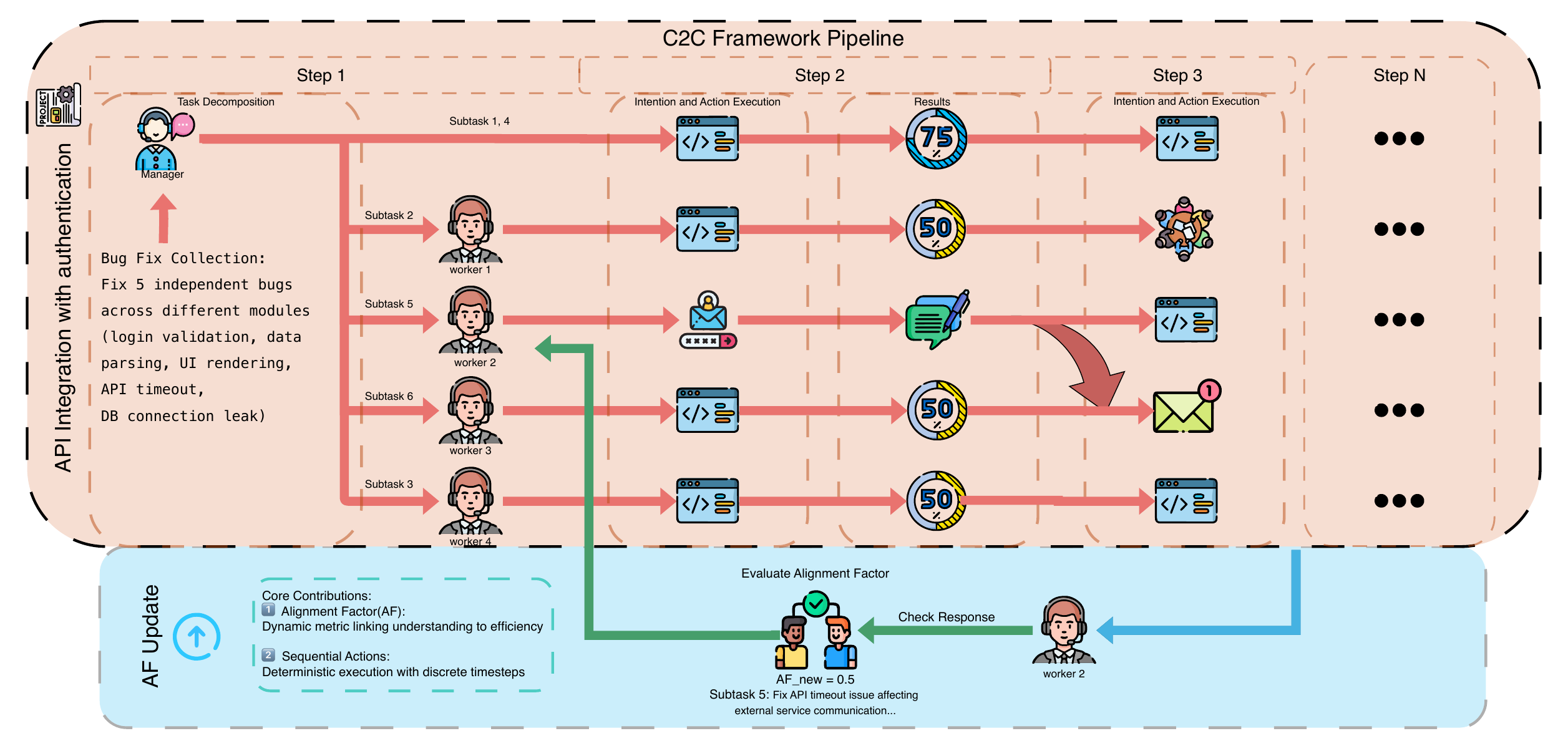}
    \caption{\textbf{C2C timeline on a software bug-fix task.} 
Step~1: the manager decomposes the task and assigns subtasks. 
Step~2: workers execute intentions and produce interim results. 
Communications update the \emph{Alignment Factor (AF)}; higher AF increases work efficiency and reshapes subsequent actions (green feedback). 
Vertical dashed dividers denote discrete timesteps in the Sequential Action Framework; the cycle repeats until completion (Step~N).}
    \label{fig:c2c_overview}
\end{figure*}


\section{Related Work}

\subsection{Multi-Agent Collaboration Frameworks}
Recent research on LLM multi-agent systems broadly spans two lines of work: task oriented execution frameworks and behavioral simulations.
Task oriented frameworks such as MetaGPT~\citep{hong2024metagpt}, ChatDev~\citep{qian2023chatdev}, and AutoGen~\citep{wu2024autogen} employ role structures, Standard Operating Procedures (SOPs), or conversable agents to solve complex engineering tasks. While effective for end-to-end task completion, these systems typically do not explicitly model communication as a constrained resource: interaction is commonly treated as low friction message passing, without representing temporal latency or cognitive overhead that shape real world teamwork.
Conversely, behavioral simulations such as Generative Agents~\citep{park2023generative} and AgentSociety~\citep{piao2025agentsociety} focus on reproducing believable human social dynamics and daily routines, but are less tied to goal directed engineering workflows. Our work bridges these directions by adopting the simulation perspective while targeting task-oriented collaboration: we explicitly model coordination constraints and costs to enable controlled study of communication strategies under resource limitations.

\subsection{Communication Costs and Constraints}
Communication overhead is a foundational concern in organizational theory and collaborative systems, and has also been widely studied in multi-agent reinforcement learning (MARL). In MARL, agents learn to selectively communicate or gate messages to reduce bandwidth usage and coordination overhead~\citep{singh2018learning,kim2019learning}. In LLM-based multi-agent systems, communication is predominantly natural language and is often treated as part of the reasoning process rather than an explicit resource. Although recent work structures agent interaction through protocols such as debate and other deliberative exchanges~\citep{du2023improving,khan2024debating}, mainstream frameworks still rarely represent costs such as asynchronous delay or context switch overhead.
Empirical studies in software engineering further suggest that coordination costs (e.g., synchronous meetings and scheduling friction) can dominate delivery time in distributed teams~\citep{mok2023challenging,yang2022effects}. Our framework explicitly models such costs via a discrete event simulation engine, distinguishing our approach from systems that view communication primarily as a byproduct of reasoning.

\subsection{Shared Mental Models and Alignment}
Effective teamwork relies on \textit{Shared Mental Models} (SMMs), the overlapping knowledge structures held by team members that support coordinated action~\citep{cannon1993shared,mathieu2000influence}. In human teams, misalignment can lead to ``process loss'', where effort is diverted to resolving misunderstandings rather than executing tasks.
For LLM agents, assessing alignment remains challenging. Most evaluations emphasize outcome correctness (e.g., pass@1 on SWE-bench~\citep{jimenez2023swe,yang2024swe}) rather than process quality. While recent benchmarks and analyses examine interaction trajectories~\citep{liu2023agentbench,yang2024swe}, they do not provide a dynamic measure of consensus during collaboration. We introduce the \textbf{Alignment Factor (AF)} as a computational proxy for SMM, enabling quantitative measurement of how communication updates shared context and how evolving alignment relates to downstream execution efficiency.

\section{C2C Framework}
\label{sec:framework}

Communication to Completion is a simulation framework for deciding \emph{when}, \emph{with whom}, and \emph{how} to communicate in multi-agent collaboration. Figure~\ref{fig:c2c_overview} provides an overview of the mechanism. We organize the discussion as follows: \S\ref{sec:saf} introduces the discrete event engine that ensures deterministic execution, \S\ref{sec:alignment} describes the Alignment Factor for quantifying task understanding, \S\ref{sec:communication} details how agents make cost-aware communication decisions, and \S\ref{sec:task_management} explains hierarchical task decomposition.

\subsection{Discrete Event Simulation Engine}
\label{sec:saf}

Concurrent multi-agent execution often introduces temporal ambiguity. To ensure reproducibility, we employ a discrete event simulation engine that constrains each agent to exactly one action per timestep, yielding deterministic state transitions. 

\paragraph{Actions}
The engine defines four actions capturing common collaborative behavior:
\emph{work} (task execution with time cost), \emph{communicate} (compose/send a message), \emph{reply} (respond to an incoming message), and \emph{meeting} (synchronous group discussion). Each action is temporally bounded and commits a single transition.

\paragraph{Synchronized Timesteps}
All agents share a fixed temporal grid. At timestep $t$, agent $i$ selects and executes $a_i^t$, then the system advances to $t{+}1$. Formally, all actions at $t$ finish before any action at $t{+}1$ begins:
\begin{equation}
\forall i,j:\; a_i^t \;\text{completes before}\; a_j^{t+1}\;\text{starts.}
\end{equation}

\paragraph{Forward-only Delivery}
Messages sent at $t$ are delivered at $t{+}1$, preventing instantaneous feedback loops and enforcing causal consistency. A communication buffer stages pending deliveries and flushes at timestep boundaries.

\subsection{Alignment Factor}
\label{sec:alignment}

The \emph{Alignment Factor} (AF) serves as a proxy for the \textbf{Shared Mental Model}~\citep{cannon1993shared}, measuring an agent's task specific understanding. 
Unlike static capability scores, AF evolves through communication. For agent $i$ on task $j$, $AF_{i,j}\in[0.01,1.00]$ denotes comprehension quality. We initialize $AF_{i,j}{=}0.3$ to reflect the initial ambiguity.

When agent $i$ receives information, the AF is updated based on the communication's quality. 
To rigorously measure this without agent bias, the update is computed by the \textbf{simulation environment} (using ground-truth context) rather than the agent itself: 
  \begin{equation}
  AF_{i,j}^{\text{new}}=\min\!\bigl(1.0,\; AF_{i,j}^{\text{old}} +
  \Delta_{\text{eval}}\bigr),
  \end{equation}
where $\Delta_{\text{eval}} \in [0, 0.5]$ is computed based on: (i) knowledge gap resolution, (ii) relevance to task requirements, and (iii) actionability. 

If agent $i$ spends $h$ hours on task $j$, the effective progress is
\begin{equation}
\text{EffectiveProgress} \;=\; h \cdot AF_{i,j},
\end{equation}
A low AF impedes progress, creating a natural incentive to seek clarification despite the communication cost.

\subsection{Agent Communication Decisions}
\label{sec:communication}

Agents autonomously decide communication strategies based on situational awareness. Crucially, this decision process is cost aware, balancing the need for alignment against simulated channel constraints.

\paragraph{Communication Initiation}
Agents may initiate communication in response to task situations such as technical difficulties (\textsc{help\_request}) or uncoordinated dependencies (\textsc{meeting\_invite}). The decision emerges from the agent's assessment of its current AF versus the time cost of interaction.

\paragraph{Recipient Selection}
Agents select recipients based on skill relevance and historical interaction, reflecting realistic collaboration patterns.

\paragraph{Channel Selection and Costs}
Agents operate on a utility maximization principle. We model the utility $U_{i,ch}$ for agent $i$ choosing channel $ch$ as:
\begin{equation}
U_{i,ch} = w_1 \cdot \mathcal{S}(D_{task}, B_{ch}) - w_2 \cdot \mathcal{C}(T_{lat}, N) + \rho_{i,ch}
\end{equation}
where:
\begin{itemize}
    \item \textbf{Task Suitability ($\mathcal{S}$):} Measures how well the channel's bandwidth $B_{ch}$ matches the current task complexity $D_{task}$. Complex tasks yield higher utility in high bandwidth channels (Meetings), whereas simple tasks (e.g., status check) favor low ones.
    \item \textbf{Time Cost ($\mathcal{C}$):} Penalizes transmission latency $T_{lat}$ and synchronization overhead for $N$ participants.
    \item \textbf{Agent Preference ($\rho_{i,ch}$):} Captures agent specific biases,  $\rho > 0$ for preferred channels.
\end{itemize}
This composite formulation ensures that agents do not merely seek the fastest route, but dynamically select the most appropriate channel given the task's nature and their role constraints.
\paragraph{Message Composition}
Communication content is generated based on the agent's current context, including task description and alignment gaps. Messages are composed to be informative.

\subsection{Hierarchical Task Management}
\label{sec:task_management}

Complex tasks are decomposed by manager agents and tracked as a directed acyclic graph (DAG) of subtasks and dependencies.

The manager performs task decomposition in the beginning of the simulation, by analyzing the task requirements and proposing several subtasks according to team size and skills. Subtasks are connected through dependency edges. The manager tracks these dependencies and coordinates task assignments to ensure workers receive suitable subtasks. And the manager updates the task graph as workers complete subtasks, monitoring overall progress. Parent task progress is computed as a weighted average of subtask progress:
  \begin{equation}
  \text{ParentProgress}=\frac{\sum_{i\in\text{subtasks}}
  w_i\,P_i}{\sum_{i\in\text{subtasks}} w_i},
  \end{equation}
where $w_i$ represents the effort estimate for subtask $i$. The parent task is marked complete only when all required subtasks reach done status, ensuring accurate project tracking.

\subsection{Intention-Based Agent Decision Making}
\label{sec:decision}

Each agent makes decisions through an intention-based mechanism that evaluates the current situation and generates contextually appropriate actions.

\paragraph{Context Formation}
At each timestep, the agent constructs a comprehensive context that includes: (i) currently assigned tasks and their completion status, (ii) alignment factors for each assignment indicating task understanding, (iii) recent communications including pending requests and received guidance,
and (iv) team state including teammate skills and availability

\begin{table*}[htb]
    \centering
    \resizebox{\textwidth}{!}{%
    \begin{tabular}{l l c c c c}
    \toprule
    \textbf{Metric} & \textbf{Complexity} & \textbf{No Comm} & \textbf{Fixed Steps} & \textbf{Free Comm}$^\dagger$ & \textbf{C2C (Ours)} \\
    \midrule
    
    \multirow{3}{*}{\textbf{Task Completion Rate} (\%)}
    & Simple & $100$ & $100$ & $96 \pm 8$ & $100$ \\
    & Medium & $100$ & $100$ & $100$ & $100$ \\
    & Complex & $100$ & $100$ & $100$ & $100$ \\
    \midrule
    
    \multirow{3}{*}{\textbf{Avg Completion Time} (hours) $\downarrow$}
    & Simple & $7.00$ & $\mathbf{5.00 \pm 0.3}$ & $6.00 \pm 0.4$ & $5.50 \pm 0.25$ \\
    & Medium & $20.0$ & $14.75 \pm 0.75$ & $14.5 \pm 1.1$ & $\mathbf{13.00 \pm 0.9}$ \\
    & Complex & $33.5$ & $36.25 \pm 3.25$ & $28.0 \pm 2.8$ & $\mathbf{24.75 \pm 1.5}$ \\
    \midrule
    
    \multirow{3}{*}{\textbf{Communication Cost} (hours) $\downarrow$}
    & Simple & -- & $2.03 \pm 0.1$ & $3.20 \pm 0.6$ & $\mathbf{1.94 \pm 0.2}$ \\
    & Medium & -- & $2.75 \pm 0.2$ & $5.40 \pm 1.5$ & $\mathbf{3.26 \pm 0.2}$ \\
    & Complex & -- & $8.12 \pm 0.9$ & $14.55 \pm 3.2$ & $\mathbf{7.02 \pm 0.8}$ \\
    \midrule
    
    \multirow{3}{*}{\textbf{Alignment Score} (AF) $\uparrow$}
    & Simple & $0.30$ & $0.55 \pm 0.05$ & $\mathbf{0.60 \pm 0.04}$ & $0.51 \pm 0.03$ \\
    & Medium & $0.30$ & $0.59 \pm 0.06$ & $0.63 \pm 0.05$ & $\mathbf{0.68 \pm 0.04}$ \\
    & Complex & $0.30$ & $0.53 \pm 0.08$ & $\mathbf{0.65 \pm 0.07}$ & $0.55 \pm 0.05$ \\
    \midrule
    
    \multirow{3}{*}{\textbf{Efficiency} (Work/Time) $\uparrow$}
    & Simple & $1.14$ & $\mathbf{1.60 \pm 0.1}$ & $1.33 \pm 0.7$ & $1.45 \pm 0.07$ \\
    & Medium & $1.20$ & $1.63 \pm 0.1$ & $1.66 \pm 0.12$ & $\mathbf{1.85 \pm 0.13}$ \\
    & Complex & $1.19$ & $1.10 \pm 0.09$ & $1.43 \pm 0.14$ & $\mathbf{1.62 \pm 0.1}$ \\
    \bottomrule
    \end{tabular}
    }
    \caption{
    Main results comparing collaboration policies on GPT-4o agents. 
    \textbf{Free Comm}$^\dagger$ represents an unconstrained baseline where agents communicate without simulated delays, resulting in faster raw completion but high communication costs.
    Values are reported as \textit{mean $\pm$ standard deviation} over 5 runs.
    \textbf{C2C} achieves the highest \textbf{Efficiency} (task progress normalized by total time investment) on Medium and Complex tasks, demonstrating the value of cost-aware strategy. 
    ($\downarrow$: lower is better, $\uparrow$: higher is better).
    }
    \label{tab:main_results}
\end{table*}

\paragraph{Intention}
Given this context, the agent uses an LLM to generate an intention. The generated intention translates into actions within the simulation:
  \begin{itemize}
  \item \textbf{Work}: Continue task execution (efficiency modulated by AF).
  \item \textbf{Help/Clarification}: Request assistance (incurs communication cost).  
  \item \textbf{Coordination}: Propose meetings (high synchronization cost).
  \item \textbf{Reporting}: Share progress.
  \end{itemize}

When generating communication intentions, the agent also determines appropriate recipients and selects suitable channels (chat, email, meeting) based on urgency and complexity.

\paragraph{Adaptive Behavior} This intention-based method enables agents to exhibit adaptive behavior without hard coded rules. The communication patterns emerge naturally from the agents' contextual reasoning rather than threshold triggers.

By integrating these components, C2C enables agents to collaborate strategically and adaptively based on actual task needs.
\section{Experiments}
\label{sec:experiments}

\subsection{Experimental Setup}
\label{sec:setup}

To evaluate the performance of C2C on realistic tasks, we constructed a benchmark dataset consisting of \textbf{15 software engineering workflows} across three complexity tiers (5 tasks per tier) to ensure robustness across different business domains (see Appendix A for full task list). For clarity in our detailed ablation studies, we report metrics from the \textbf{representative task} of each tier:
\begin{itemize}
    \item \textbf{Simple} (8 hours): Basic SWE operations (e.g., bug fixes, refactoring).
    \item \textbf{Medium} (24 hours): Module integration (e.g., API integration with authentication).
    \item \textbf{Complex} (40 hours): Full-stack system design (e.g., user authentication service).
\end{itemize}
Each task requires diverse skills including front-end development, back-end systems, database management, and testing. Tasks are decomposed into subtasks with explicit dependencies, mirroring real-world software development scenarios.

\begin{figure*}[ht!]
    \centering
    \begin{subfigure}[b]{0.325\textwidth}
        \centering
        \includegraphics[width=\linewidth]{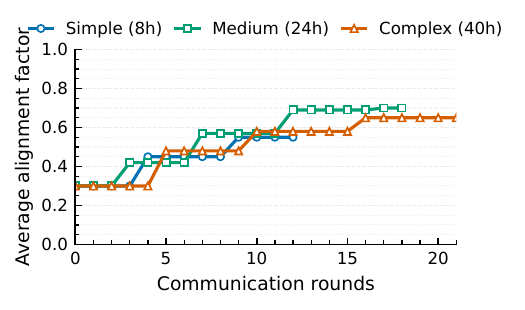}
        \caption{Alignment evolution over rounds}
        \label{fig:alignment_evolution_a}
    \end{subfigure}
    \begin{subfigure}[b]{0.325\textwidth}
        \centering
        \includegraphics[width=\linewidth]{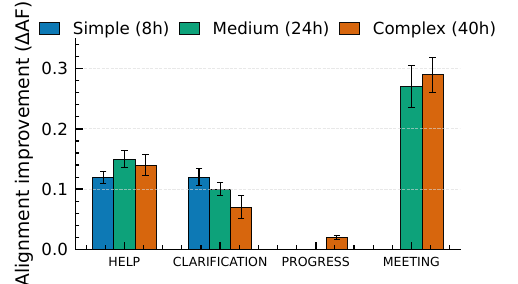}
        \caption{Impact of communication type}
        \label{fig:alignment_evolution_b}
    \end{subfigure}
    \begin{subfigure}[b]{0.325\textwidth}
        \centering
        \includegraphics[width=\linewidth]{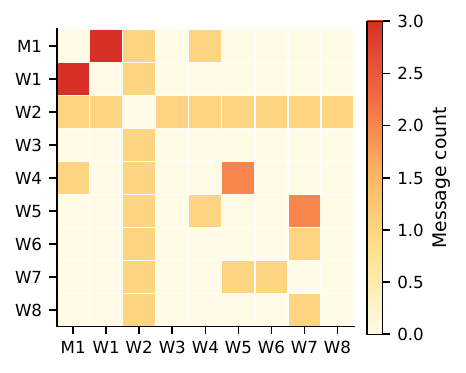}
        \caption{Communication heatmap}
        \label{fig:alignment_evolution_d}
    \end{subfigure}
    
    \caption{
        Analysis of the Alignment Factor (AF) dynamics. 
        (a) Average alignment improves over successive communication rounds across all task complexities. 
        (b) High intent communications like MEETINGS and HELP requests yield the largest gains in alignment. 
        (c) The communication heatmap reveals a manager centric coordination pattern, with the manager (M1) acting as the central hub.
    }
    \label{fig:alignment_evolution}
\end{figure*}

\paragraph{Team Configurations} We systematically vary team sizes from 5 to 17 agents: 1M+4W (5 agents), 1M+8W (9 agents), and 1M+16W (17 agents), where M denotes manager and W denotes worker. Additionally, we evaluate multi-task scenarios with 1M+8W handling 2 concurrent tasks.

\paragraph{Models and Implementation} Agents are powered by \textbf{GPT-4o} (temperature 0.7). We also evaluate \textbf{Claude 3.5 Sonnet} to test generalization. The discrete event simulation engine operates with 0.25-hour time steps over a maximum of 160 steps. Communication costs model real world constraints: Email (9 min delay + drafting), Chat (3 min), and Meetings (30 min blocking cost).

\paragraph{Baselines} We compare C2C against three baselines: 
(1) \textbf{No Communication}: Agents work independently.
(2) \textbf{Fixed Steps}: Agents communicate on a rigid schedule (every 16 steps).
(3) \textbf{Free Communication}: An unconstrained baseline where agents communicate without simulated latency, representing idealized interaction.

\paragraph{Metrics} We report: \textbf{task completion rate} (percentage of tasks completed within time budget), \textbf{average completion time} (hours to finish successful tasks), \textbf{communication cost} (total time spent on communication activities), \textbf{alignment score} (average agent task alignment factor), and \textbf{efficiency} (ratio of productive work to total time investment). Each configuration is evaluated across all three task complexity tiers.

\subsection{Main Results}
\label{sec:main_results}

Table~\ref{tab:main_results} presents the performance comparison. While all methods achieve high completion rates, C2C consistently demonstrates superior \textbf{Efficiency} on Medium (1.85) and Complex (1.62) tasks. Comparison with the \textbf{Free Communication} baseline reveals the impact of communication overhead. Although Free Comm allows agents to communicate instantly, this leads to excessive interaction (e.g., 14.55 hours cost on Complex tasks vs. 7.02 hours for C2C). This chatter distracts agents from execution, effectively slowing down the overall progress. C2C, by weighing the cost of communication against alignment gains, finds an optimal balance, engaging in just enough coordination to maintain high Alignment Factor (AF) without incurring high costs.

To understand the dynamics behind this efficiency, \textbf{Figure~\ref{fig:alignment_evolution}} analyzes the evolution of the Alignment Factor. As shown in \textbf{Figure~\ref{fig:alignment_evolution}(a)}, alignment scores start low and improve progressively. \textbf{Figure~\ref{fig:alignment_evolution}(b)} highlights that high cost interactions like \textbf{Meetings} yield the largest alignment gains
justifying their use for critical synchronization, whereas low cost messages provide smaller increments. The heatmap in \textbf{Figure~\ref{fig:alignment_evolution}(c)} confirms a manager coordination pattern, where the manager acts as a hub to distribute context efficiently.

\subsection{Generalization across Models}
\label{sec:generalization}

To verify that our findings are not specific to GPT-4o, we evaluated C2C using three frontier models: \textbf{GPT-5.2}, \textbf{Claude Sonnet 4.5}, and \textbf{Gemini 2.5 Pro}.
As shown in Table~\ref{tab:model_generalization}, the performance advantage of C2C remains consistent across all backbones.
The \textbf{Claude Sonnet 4.5} model, known for its specialized coding capabilities, achieves the highest raw performance. However, without constraints, it still incurs significant communication overhead in the \textit{Free Communication} baseline (Efficiency 1.96).
Applying the C2C framework boosts its efficiency to \textbf{2.40} (a \textbf{22\%} relative improvement).
Similarly, \textbf{GPT-5.2} sees an efficiency gain from 2.08 to \textbf{2.22}, and \textbf{Gemini 2.5 Pro} from 1.88 to \textbf{2.09}.
These results confirm that while stronger models can execute tasks faster, they are not immune to coordination friction. C2C provides a universal optimization layer that scales with model capability, ensuring that high performance agents focus on execution rather than redundant communication.

\begin{table}[hb]
    \centering
    \resizebox{\columnwidth}{!}{%
    \begin{tabular}{l l c c c}
    \toprule
    \textbf{Backbone Model} & \textbf{Policy} & \textbf{Time (h)} $\downarrow$ & \textbf{Comm Cost (h)} $\downarrow$ & \textbf{Efficiency} $\uparrow$ \\
    \midrule
    
    \multirow{2}{*}{\textbf{GPT-5.2}} 
    & Free Comm$^\dagger$ & $11.50$ & $4.20$ & $2.08$ \\
    & \textbf{C2C (Ours)} & $\mathbf{10.80}$ & $\mathbf{2.15}$ & $\mathbf{2.22}$ \\
    \midrule
    
    \multirow{2}{*}{\textbf{Claude Sonnet 4.5}} 
    & Free Comm$^\dagger$ & $12.25$ & $4.80$ & $1.96$ \\
    & \textbf{C2C (Ours)} & $\mathbf{10.00}$ & $\mathbf{2.40}$ & $\mathbf{2.40}$ \\
    \midrule

    \multirow{2}{*}{\textbf{Gemini 2.5 Pro}} 
    & Free Comm$^\dagger$ & $12.75$ & $5.10$ & $1.88$ \\
    & \textbf{C2C (Ours)} & $\mathbf{11.50}$ & $\mathbf{2.65}$ & $\mathbf{2.09}$ \\
    
    \bottomrule
    \end{tabular}
    }
    \caption{
    Generalization analysis on \textbf{Medium} complexity tasks. 
    Even with highly capable models, the unconstrained \textbf{Free Comm} baseline suffers from communication overhead. 
    \textbf{C2C} consistently improves \textbf{Efficiency} by minimizing coordination costs, demonstrating that the framework's value scales with model capability.
    }
    \label{tab:model_generalization}
\end{table}

\subsection{Scalability and Multi-Task Analysis}
\label{sec:scalability}

Table~\ref{tab:scalability} demonstrates C2C's effectiveness across team sizes. The framework exhibits sub-linear scaling in communication cost: while team size increases 3.4$\times$ (5 to 17 agents), communication cost only increases 86\%. Speedup analysis reveals that performance improves with team size, achieving $\mathbf{1.95 \times}$ speedup with 17 agents.
In multi-task scenarios (1M+8W, 2 tasks), C2C evolves hub-and-spoke topologies to manage context switching, keeping completion time (21h) significantly below linear scaling assumptions.

\setlength{\tabcolsep}{2pt}
\begin{table}[h]   
    \centering
    \resizebox{\columnwidth}{!}{%
    \begin{tabular}{lcccc}
    \toprule
    \textbf{Configuration} & \textbf{Completion Time} & \textbf{Comm/Agent} & \textbf{Comm Cost} & \textbf{Speedup} \\
    \midrule
    1M + 4W & 13h & 3.1 & 2.75 & 1.54 \\
    1M + 8W & 11.25h & 2.1 & 3.78 & 1.78 \\
    1M + 16W & 10.25h & 2.6 & 5.12 & \textbf{1.95} \\
    \midrule
    1M+8W (2 tasks) & 21h & 2.3 & 4.64 & 1.35 \\
    \bottomrule
    \end{tabular}
    }
    \caption{Scalability and multi-task analysis. Speedup is relative to the "No Communication" baseline.}
    \label{tab:scalability}
\end{table}

Detailed analysis reveals that the incremental gain in speedup lessens as the team grows (from 1.54$\times$ to 1.95$\times$). This suggests that simply adding more workers yields progressively smaller advantages due to the inherent coordination overhead required to synchronize a larger workforce, aligning with Brooks' Law in software engineering.

The multi-task evaluation with 1M+8W reveals C2C's ability to handle concurrent workloads effectively. When processing 2 tasks simultaneously, completion time increases from 11.25h to 21h (an 87\% increase), significantly better than naive linear scaling. Analysis of communication patterns shows that C2C naturally evolves hub-and-spoke topologies with managers as primary coordinators, avoiding the quadratic communication complexity that plagues peer-to-peer approaches. In multi-task scenarios, agents exhibit sophisticated context switching behavior, maintaining separate alignment factors per task and prioritizing communications based on overall workflow optimization.

\begin{figure*}[ht!]
    \centering
    \begin{subfigure}[b]{0.47\textwidth}
        \centering
        \includegraphics[width=\linewidth]{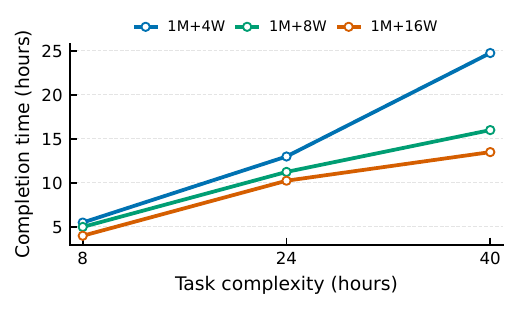}
        \caption{Time vs. Complexity.}
        \label{fig:task_complexity_a}
    \end{subfigure}
    \hfill
    \begin{subfigure}[b]{0.47\textwidth}
        \centering
        \includegraphics[width=\linewidth]{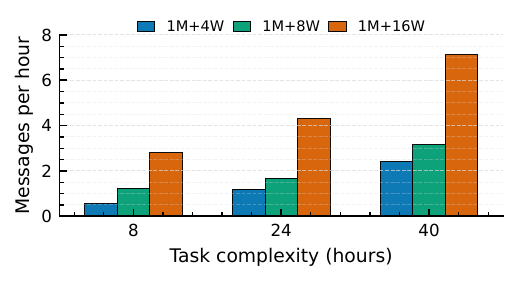}
        \caption{Communication Freq. vs. Complexity.}
        \label{fig:task_complexity_b}
    \end{subfigure}

    \vspace{1em}

    \begin{subfigure}[b]{0.47\textwidth}
        \centering
        \includegraphics[width=\linewidth]{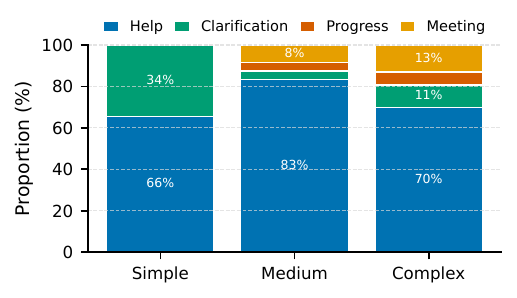}
        \caption{Message Type Distribution.}
        \label{fig:task_complexity_c}
    \end{subfigure}
    \hfill
    \begin{subfigure}[b]{0.47\textwidth}
        \centering
        \includegraphics[width=\linewidth]{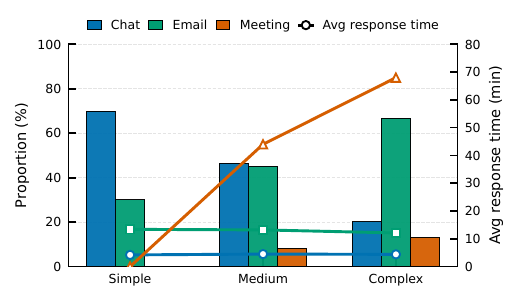}
        \caption{Communication Method Analysis.}
        \label{fig:task_complexity_d}
    \end{subfigure}

    \caption{
    Effect of task complexity under C2C (8/24/40 hours) and team configuration
    (1M{+}4W, 1M{+}8W, 1M{+}16W).
    (a) Completion time increases with task complexity; larger worker pools shorten time with diminishing returns.
    (b) Messages per hour rise with both complexity and team size.
    (c) Help messages dominate across settings; clarification and meeting shares grow with complexity, and progress updates appear from medium upward.
    (d) Communication channels shift from chat (simple) toward email (complex), with a modest increase in meetings.
    }
    \label{fig:task_complexity_impact}
\end{figure*}

\subsection{Message Type and Decomposition}
\label{sec:message_analysis}

\paragraph{Message Type Analysis}
Table~\ref{tab:message_analysis} confirms that \textbf{MEETING\_INVITE} messages provide the highest alignment gains (+0.27), justifying their high cost for critical synchronization. \textbf{HELP\_REQUEST} follows (+0.15), while \textbf{PROGRESS\_UPDATE} has negligible immediate impact on AF but maintains long-term context.

\begin{table}[!htbp]
    \centering
    \resizebox{\columnwidth}{!}{%
    \begin{tabular}{lccc}
    \toprule
    \textbf{Message Type} & \textbf{Frequency} & \textbf{Avg Response Step} & \textbf{Impact on Alignment} \\
    \midrule
    HELP\_REQUEST & \textbf{9} & 3 & +0.15 \\
    CLARIFICATION & 0.3 & 2 & +0.10 \\
    PROGRESS\_UPDATE & 0.3 & 5 & 0 \\
    MEETING\_INVITE & 1 & 7 & +0.27 \\
    RESPONSE & 8 & -- & -- \\
    \bottomrule
    \end{tabular}
    }
    \caption{Analysis of message types showing frequency and impact on AF.}
    \label{tab:message_analysis}
\end{table}

\paragraph{Task Decomposition Quality}
As shown in Table~\ref{tab:decomposition_quality}, C2C's adaptive decomposition strategy achieves higher subtask clarity (0.95) compared to naive LLM prompting (0.72), approaching human manual quality (1.00).

\begin{table}[!htbp]
    \centering
    \resizebox{\columnwidth}{!}{%
    \begin{tabular}{lcc}
    \toprule
    \textbf{Decomposition Method} & \textbf{Subtask Clarity} & \textbf{Alignment Factor} \\
    \midrule
    Manual & 1.00 & 0.70 \\
    LLM-naive & 0.72 & 0.58 \\
    Hierarchical & 0.89 & 0.64 \\
    \textbf{C2C Adaptive} & \textbf{0.95} & \textbf{0.68} \\
    \bottomrule
    \end{tabular}
    }
    \caption{Comparison of task decomposition quality judged by LLM.}
    \label{tab:decomposition_quality}
\end{table}

\subsection{Communication Pattern Analysis}
\label{sec:comm_analysis}

Figure~\ref{fig:task_complexity_impact} shows how communication varies with task complexity and team size (1M{+}4W, 1M{+}8W, 1M{+}16W).
\textbf{(a)} Completion time increases with task complexity; adding workers shortens time but with diminishing returns.
\textbf{(b)} Message intensity (messages/hour) rises with complexity and team size.

\paragraph{Message Types} In \textbf{(c)}, simple tasks are dominated by \emph{help} and \emph{clarification} (about 66\% and 34\%, respectively). At medium complexity, \emph{help} remains the majority ($\approx$83\%), while \emph{meetings} appear ($\approx$8\%). For complex tasks, the mix diversifies: \emph{help} $\approx$70\%, and \emph{meetings} $\approx$13\% become dominant. C2C allocates most messages to high value help requests and escalates to meetings only when the expected coordination benefit exceeds the cost.

\paragraph{Channels and Latency} Panel~\textbf{(d)} indicates a shift from chat toward email as complexity grows, with a modest rise in meetings. The per channel response times for \emph{chat} and \emph{email} stay roughly flat across complexity levels (reduce slightly), whereas \emph{meeting} latency increases for complex tasks.

These patterns align with the logic of C2C: as tasks become more complex, agents continue to seek help most of the time but increasingly use meetings, while clarification needs drop. The engine selects higher yield (though costlier) channels when needed, while larger teams reduce completion time without eliminating the upward pressure on communication volume with complexity.

 

\section{Conclusion}
\label{sec:conclusion}

In this paper, we investigate the modeling of collaboration efficiency and communication constraints in multi-agent systems. Rather than focusing solely on agent architecture, we propose a multi—agent framework, Communication to Completion (C2C), to quantify the interplay between information alignment and coordination costs. By formalizing task understanding through the Alignment Factor and modeling communication as a scarce resource with explicit latency, we establish a rigorous methodology for quantifying the trade-offs between coordination overhead and execution efficiency in collaborative workplace.

Our experiments validate this modeling approach, revealing that efficiency is not merely a function of individual agent capability, but a dynamic equilibrium between alignment gains and communication overhead. The simulation results demonstrate that imposing realistic constraints drives agents to emerge with strategic behaviors, such as batching questions or escalating to meetings, which are absent in unconstrained environments. These patterns hold consistent across different backbone models , suggesting that our cost-alignment model captures fundamental dynamics of collaborative intelligence.
\paragraph{Limitations}
While the C2C framework demonstrates significant efficiency gains, we acknowledge several limitations that offer avenues for future research. First, the Alignment Factor acts as a proxy for understanding but does not guarantee functional correctness (e.g., passing unit tests). Second, our deterministic cost models simplify real world stochastic interruptions. Finally, our experiments are specific to software engineering; generalizing these dynamics to other collaborative domains remains future work.

\nocite{*}
\bibliography{arr}

\appendix

\section{Tasks Details}
\label{sec:tasks}

This part provide example task prompts used in our simulations. Each prompt specifies the description, time budget, and required skills.
\begin{figure}[h]
\begin{promptbox}{\fontppl Simple Task}
\begin{lstlisting}
description: "Fix five independent bugs across modules: login validation, data parsing, UI rendering glitches, API timeout handling, and a database connection leak. No cross-dependencies."
hours: 8.0
skills: ["backend", "frontend", "database", "api", "testing"]
\end{lstlisting}
\end{promptbox}

\begin{promptbox}{\fontppl Medium Task}
\begin{lstlisting}
description: "Integrate an external API with authentication, including token management and error/latency handling; deliver minimal usage docs."
hours: 24.0
skills: ["backend", "api", "authentication", "oauth", "testing", "documentation"]
\end{lstlisting}
\end{promptbox}

\begin{promptbox}{\fontppl Hard Task}
\begin{lstlisting}
description: "Build a user authentication service covering registration, login, password reset, OAuth 2.0 sign-in, JWT issuance/refresh, session management, and security hardening."
hours: 40.0
skills: ["backend", "security", "database", "oauth", "authentication", "frontend", "testing"]
\end{lstlisting}
\end{promptbox}
\end{figure}

\section{Implementation Details}
\label{sec:implentation}

We provide prompts used in our experiments in this section.

\begin{figure*}[htbp]
\centering
\begin{footnotesize}
\begin{minipage}{\textwidth}

\begin{promptbox}{\fontppl Decompose a task into subtasks}

Decompose this task into subtasks:\\[1em]

Task: \{task.description\}\\
Estimated hours: \{task.estimated\_hours\}\\
Required skills: \{', '.join(task.required\_skills)\}\\[1em]

Team members:\\
\{self.\_format\_team(team)\}\\
\{manager\_note\}\\
Create \{teamsize\} subtasks that:\\
1. Can be worked on independently or with minimal dependencies\\
2. Match team members' skills\\
3. Are reasonably sized (total hours should be close to the original task's estimated hours)\\
4. Cover all aspects of the original task \\[1em]

\begin{lstlisting}
Return JSON:
{{
    "subtasks": [
       {{
            "description": "Clear description of what needs to be done",
            "estimated_hours": number,
            "required_skills": ["skill1", "skill2"],
            "suggested_assignee": "team member_name (can be any team member including Manager)",
            "dependencies": []  // indices of subtasks this depends on
        }}
    ],
    "decomposition_rationale": "Brief explanation of your decomposition strategy"
}}
\end{lstlisting}
\end{promptbox}
\caption{Prompt for the manager to decompose a give task.}

\begin{promptbox}{\fontppl Intention generation}
\begin{lstlisting}
You are {context.agent.name}, a {context.agent.role.value}.
        
Current situation:
Tasks:
{self._format_tasks_for_prompt(context.current_tasks[:3], context)}
Note: Alignment affects work efficiency.(max: 1.0)
{message_info}
{last_action_info}

Analyze your situation and choose your primary intention for this step:

1. CONTINUE_TASK - Continue working on current tasks
2. CHECK_MESSAGES - Check and potentially respond to messages
3. REQUEST_HELP - Ask for other agents' help
4. NEED_CLARIFICATION - Need clarification on task requirements
5. REPORT_PROGRESS - Report progress to manager
6. SCHEDULE_MEETING - Schedule a meeting

IMPORTANT Decision Factors:
- If has MEETING_START: Almost always CHECK_MESSAGES (meeting is starting!)
- If has MEETING_INVITE: Strongly consider CHECK_MESSAGES (need to RSVP)
- If stuck on task for long: Consider REQUEST_HELP or NEED_CLARIFICATION
- Balance responsiveness with productivity

Return JSON: {{"intention": "INTENTION_NAME", "reasoning": "explanation"}}
\end{lstlisting}
\end{promptbox}
\caption{Prompt agents use to generate their intentions at each step.}

\end{minipage}
\end{footnotesize}
\vspace{-0.1in}

\end{figure*}

\begin{figure*}[htbp]
\centering
\begin{footnotesize}
\begin{minipage}{\textwidth}

\begin{promptbox}{\fontppl Decompose a task into subtasks}
\begin{lstlisting}
You are evaluating how much a received message helps an worker understand their task better.

Task Information:
- Task ID: {task_id}
- Description: {task.description}
- Required Skills: {', '.join(task.required_skills) if task.required_skills else 'Not specified'}
- Current Progress: {task.actual_hours:.1f}/{task.estimated_hours:.1f} hours
- Current Alignment Factor: {current_alignment:.2f} ({current_alignment*100:.0f}% efficiency)

Message Type: {message.message_type.value if hasattr(message, 'message_type') else 'Unknown'}
From Agent: {message.from_agent_id}

{'Original Request (sent by me):' if original_request else 'Context:'}
{original_request.content if original_request else 'This is a proactive message or the original request is not available.'}

Reply Received:
{message.content}

Alignment Factor Guidelines:
- Range: 0.01 (1% efficiency) to 1.0 (100% efficiency)
- Current value: {current_alignment:.2f}
- Alignment can increase OR decrease based on communication quality
- Clear and helpful communication should improve understanding
- Confusing, contradictory, or misleading information may reduce understanding
- Consider the overall impact on task clarity and execution confidence

Consider:
1. How directly the reply addresses the original request/confusion
2. How actionable and specific the information is
3. Whether critical blockers were resolved
4. The completeness of the response
5. Diminishing returns (harder to improve as alignment approaches 1.0)

Return a JSON response:
{{
    "new_alignment_factor": <float between 0.01 and 1.0>,
    "change": <float, positive for increase, negative for decrease>,
    "reasoning": "<brief explanation of why this change in understanding>"
}}
\end{lstlisting}
\end{promptbox}
\caption{Prompt for updating alignment factor.}

\end{minipage}
\end{footnotesize}
\vspace{-0.1in}

\end{figure*}

\end{document}